\documentclass[superscriptaddres,notitlepage]{revtex4-1}

\usepackage[margin=1.0in]{geometry}
\usepackage{verbatim}
\usepackage{graphicx}
\usepackage{amsmath}
\usepackage{epstopdf}
\usepackage{color}
\usepackage{setspace}

\begin{document}

\title{Limits of Femtosecond Fiber Amplification by Parabolic Pre-Shaping}

\author{Walter Fu,$^{1,*}$
Yuxing Tang,$^{1}$
Timothy S. McComb,$^{2,3}$
Tyson L. Lowder$^{2}$
Frank W. Wise$^{1}$
}
\affiliation{
	$^{1}$School of Applied and Engineering Physics, Cornell University, Ithaca, New York 14853, USA
}
\affiliation{
	$^{2}$nLIGHT Corporation, 5408 NE 88th Street, Bldg. E, Vancouver, Washington 98665, USA
}
\affiliation{
	$^{3}$Current affiliation: Powerlase Photonics Inc., 3251 Progress Drive, Suite 136, Orlando, Florida 32826, USA
}
\affiliation{
	$^{*}$Corresponding author: wpf32@cornell.edu
}

\begin{abstract}
We explore parabolic pre-shaping as a means of generating and amplifying ultrashort pulses.  We develop a theoretical framework for modeling the technique and use its conclusions to design a femtosecond fiber amplifier.  Starting from 9 ps pulses, we obtain 4.3 $\mu$J, nearly transform-limited pulses 275 fs in duration, simultaneously achieving over 40 dB gain and 33-fold compression.  Finally, we show that this amplification scheme is limited by Raman scattering, and outline a method by which the pulse duration and energy may be further improved and tailored for a given application.
\end{abstract}

\maketitle

\section{Introduction}

In recent years, the rapid growth of nonlinear optical techniques has spurred the demand for ultrafast, microjoule-level, pulsed sources.  While solid-state systems readily reach the desired performance level, many applications benefit from the compact and robust nature of fiber devices.  Such systems can be spliced to provide alignment-free behavior, while boasting good spatial properties even at large average powers.  However, high-energy, short-pulse amplification in fiber poses its own set of problems.  In this regime, complex interactions between chromatic dispersion and fiber nonlinearities conspire to reduce the quality and peak power of the amplified pulses.

A number of techniques exist for managing nonlinearity in high-energy fiber amplifiers. Linear chirped-pulse amplification can easily reach energies in the tens of microjoules without requiring rod amplifiers or auxiliary techniques, with pulse durations of only a couple hundred femtoseconds \cite{Strickland1985,Roser2005}. This approach relies on reversible temporal stretching to reduce the accumulated nonlinear phase, quantified by the B-integral $B(z) \equiv \int_0^L \mathrm{d}z~\gamma(z) P(z)$ for nonlinear coefficient $\gamma$ and peak power $P$, to less than $\pi$.  However, the use of fiber stretchers results in longer pulses due to uncompensated third-order dispersion, while bulk stretchers undercut many of the benefits of fiber systems.  Self-similar amplifiers can handle high nonlinearity and produce sub-100-fs pulses in a fiber-integrated format, but their rapid spectral broadening coupled with the finite gain bandwidth typically limits them to $\approx$1 $\mu$J \cite{Fermann2000,Chang2004}. Cubicon amplifiers are scalable to 100 $\mu$J and $B \gtrsim 4\pi$, but they require specially-engineered input spectra, and reported pulse durations exceed 400 fs \cite{Shah2005, Zeludevicius2013}. While nonlinear chirped-pulse amplification has been demonstrated at 30 $\mu$J and 240 fs with $B \approx 18 \pi$, its performance drops noticeably if the balance between nonlinearity and third-order-dispersion is disturbed \cite{Kuznetsova2007}.  Numerous other results exhibit very promising performance without requiring rod-type amplifiers \cite{Hanna2009,Zhao2014}, even reaching the millijoule energy level \cite{Galvanauskas2001}, but are somewhat limited by their reliance on grating stretchers.

Parabolic pre-shaping was recently proposed by Pierrot and Salin as a new, stretcher-free amplification technique capable of producing high-quality, microjoule-scale pulses \cite{Pierrot2013}.  In its first demonstration, 27 ps seed pulses were amplified, reaching 49 $\mu$J after compression near the transform limit of 780 fs.  Not only did the pulse experience a 35-fold compression with minimal loss of quality, but it also accumulated $B=22\pi$, making this approach one of the most nonlinearity-tolerant of any amplification scheme.  In addition to its potential as an ultrafast amplifier, this system was notable for its ability to generate high-quality, sub-picosecond pulses from a 27-ps oscillator.  This feature may be attractive for applications requiring synchronized broadband and narrowband pulses, such as multimodal imaging sources for coherent anti-Stokes Raman scattering microscopy \cite{Evans2008}.  Despite these impressive results, the pulses obtained by Pierrot and Salin remain too long for many applications, and the paper does not address the limitations of their system.  A natural question to ask is what the limiting factors might ultimately be, and whether they permit extending parabolic pre-shaping to even shorter pulses.  In particular, we are interested in the possibility of generating 100-fs pulses at microjoule energies, due to the wealth of applications such as nonlinear microscopy.

Here, we assess parabolic pre-shaping from a theoretical standpoint and explore the extent of its performance.  We develop a simple, analytical model of the technique and use it to predict how the output pulses vary with the system parameters.  Using this information, we demonstrate an amplifier based on parabolic pre-shaping that is consistent with the theoretical limit we predict.  Starting from sub-nanojoule, 9 ps seeds, our system generates nearly transform-limited pulses shorter than 300 fs with energies exceeding 4 $\mu$J, an attractive region of parameter space for many applications.  We confirm that parabolic pre-shaping is ultimately limited by stimulated Raman scattering, and propose a route by which even shorter pulses that would be truly competitive with established techniques might be obtained from a similar system.

\section{Theoretical Framework}

\subsection{Analytical Modeling}

\begin{figure}[htb]
\centering
{\includegraphics[width=0.7\linewidth]{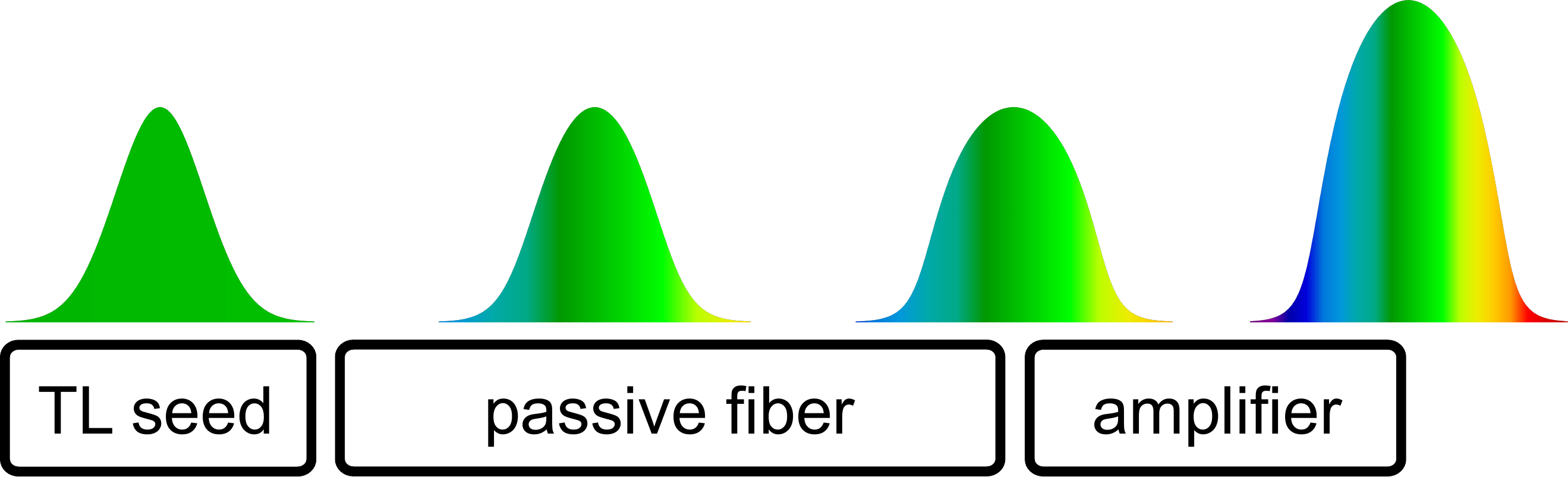}}
\caption{Illustration of an archetypal parabolic pre-shaping system and the pulse's condition at various stages.}
\label{fig:PPScartoon}
\end{figure}

The concept underlying parabolic pre-shaping is illustrated in Figure \ref{fig:PPScartoon}.  A nearly transform-limited (TL) pulse propagating in passive, normal-dispersion fiber nonlinearly reshapes itself, transiently becoming parabolic without significantly broadening in duration \cite{Finot2007}.  Amplifying the pulse in this form causes it to accumulate an essentially linear chirp due to self-phase modulation, and the resulting pulse can be dechirped using a standard grating compressor.  Although pre-shaping a pulse to reduce deleterious nonlinearities is not a new idea, previous works relied on particular input pulses \cite{Schreiber2007} or active pulse shaping \cite{Schimpf2007}, while parabolic pre-shaping benefits from requiring only a well-behaved seed pulse and a passive fiber.  We furthermore emphasize that this process is distinct from self-similar amplification, where the pulse duration, bandwidth, and energy are all constrained to increase in concert with one another and the pulse is limited by the edges of the gain spectrum \cite{Fermann2000}.  By contrast, parabolic pre-shaping decouples these quantities: the temporal evolution remains essentially static, while the energy is free to grow much larger before the finite gain bandwidth becomes problematic.

We first explore the limits of parabolic pre-shaping using a simple, analytical approach.  We model the pulse in the gain fiber as a static, linearly-chirped parabola, neglecting chromatic dispersion, gain dispersion, and higher-order nonlinear effects.  Under the influence of broadband gain and self-phase modulation, the complex pulse envelope $A$ can be written \textit{modulo} constant factors as:

\begin{equation}
A(z,t) = e^{gz/2}~\sqrt{1 - \frac{t^2}{\tau^2}}~e^{-i(B+B_0)t^2/\tau^2},~~~|t|<\tau
\end{equation}

\noindent Here, $g$ is the differential gain coefficient, $B=B(z)$ is the B-integral for the amplifier, and $B_0$ is an empirical factor which accounts for the pulse chirp at the entrance of the gain fiber but does not necessarily correspond to the actual B-integral for the passive shaping process.  The full-width-half-maximum (FWHM) of the parabolic pulse is $\mathrm{FWHM_0} = \tau \sqrt{2}$.  Using the method of stationary phase in the limit of large chirp, we write the Fourier transform $\hat{A}(z,\omega)$ as:

\begin{eqnarray}
\hat{A}(z,\omega) &\equiv& \int_{-\infty}^{\infty} \mathrm{d}t~A(z,t) e^{i \omega t} \\
&\sim& e^{gz/2}~\sqrt{1 - \frac{\omega^2}{\Omega^2}}~e^{i\tau\omega^2/(2\Omega)}
\end{eqnarray}

\noindent where $\Omega \equiv 2[B(z)+B_0]/\tau$ encapsulates the z-dependence, and we have again dropped the constant factors.  Analytically taking the inverse transform of $|\hat{A}(z,\omega)|$ then gives the transform-limited pulse envelope in terms of $J_n(x)$, the Bessel function of the first kind:

\begin{equation}
A_\mathrm{TL}(z,t) \propto e^{gz/2}~\frac{J_1(\Omega t)}{\Omega t}
\end{equation}

\noindent This allows us to calculate the transform-limited duration, $\mathrm{FWHM_{TL}}$, in terms of the input duration and the nonlinear phase accumulation:

\begin{equation}
\mathrm{FWHM_{TL}} \approx 1.14~\frac{\mathrm{FWHM_0}}{B+B_0}
\label{eq:TL}
\end{equation}

\noindent Equation~\ref{eq:TL} is the first main result of this analysis.  Because the ideal pulse modeled here has a linear chirp, we assume that it can be dechirped close to the transform limit, making $\mathrm{FWHM_{TL}}$ a reasonable proxy for the actual pulse duration after dechirping.  We see that, under fixed $B$, the output pulse duration scales linearly with the seed duration, which immediately suggests a means of scaling to shorter pulses.  Increasing $B$ has a similar effect, but risks introducing other, adverse, phenomena.

As $B$ becomes very large, we expect Raman scattering in particular to become limiting.  Although Raman scattering is generally modeled as a nonlinear convolution, it becomes analytically tractable if we neglect dispersive effects.  This simplification corresponds to the limit of long pulses and short gain fibers, such that group-velocity mismatch is insignificant.  In this limit, we can apply the quasi-CW approximation and calculate the growth of the Stokes wave $P_S$ due to the strong pump pulse $P_P$ on a point-by-point basis in the time domain:

\begin{equation}
\frac{\mathrm{d}P_S}{\mathrm{d}z} \sim \frac{g_R}{A_\mathrm{eff}} P_P P_S
\label{eq:SRS_difeq}
\end{equation}

\noindent Following Smith \cite{Smith1972}, we solve equation \ref{eq:SRS_difeq} to obtain our second main result:

\begin{equation}
P_S(z) \sim \sqrt{\pi}~h \nu_S \Delta \nu~\frac{e^x}{\sqrt{x}},~~~x \equiv \frac{\lambda_P g_R}{2 \pi n_2} B(z) \gg 1
\label{eq:SRS_solution}
\end{equation}

\noindent Note that, due to the approximations made and the exponential dependences, equation \ref{eq:SRS_solution} should not be interpreted quantitatively.  Rather, our focus lies in the fact that, amidst the clusters of constants, the Stokes wave depends only on $B$.  It follows that if there is a maximum tolerable Stokes power, there must also be a maximum $B$, and that this limit is insensitive to the parameters of the narrowband seed pulse.  This furthermore precludes arbitrarily scaling equation \ref{eq:TL} via its denominator, leaving the seed duration as the sole means of realizing shorter output pulses.

\subsection{Numerical Trends}

\begin{figure}[htb]
\centering
{\includegraphics[width=0.5\linewidth]{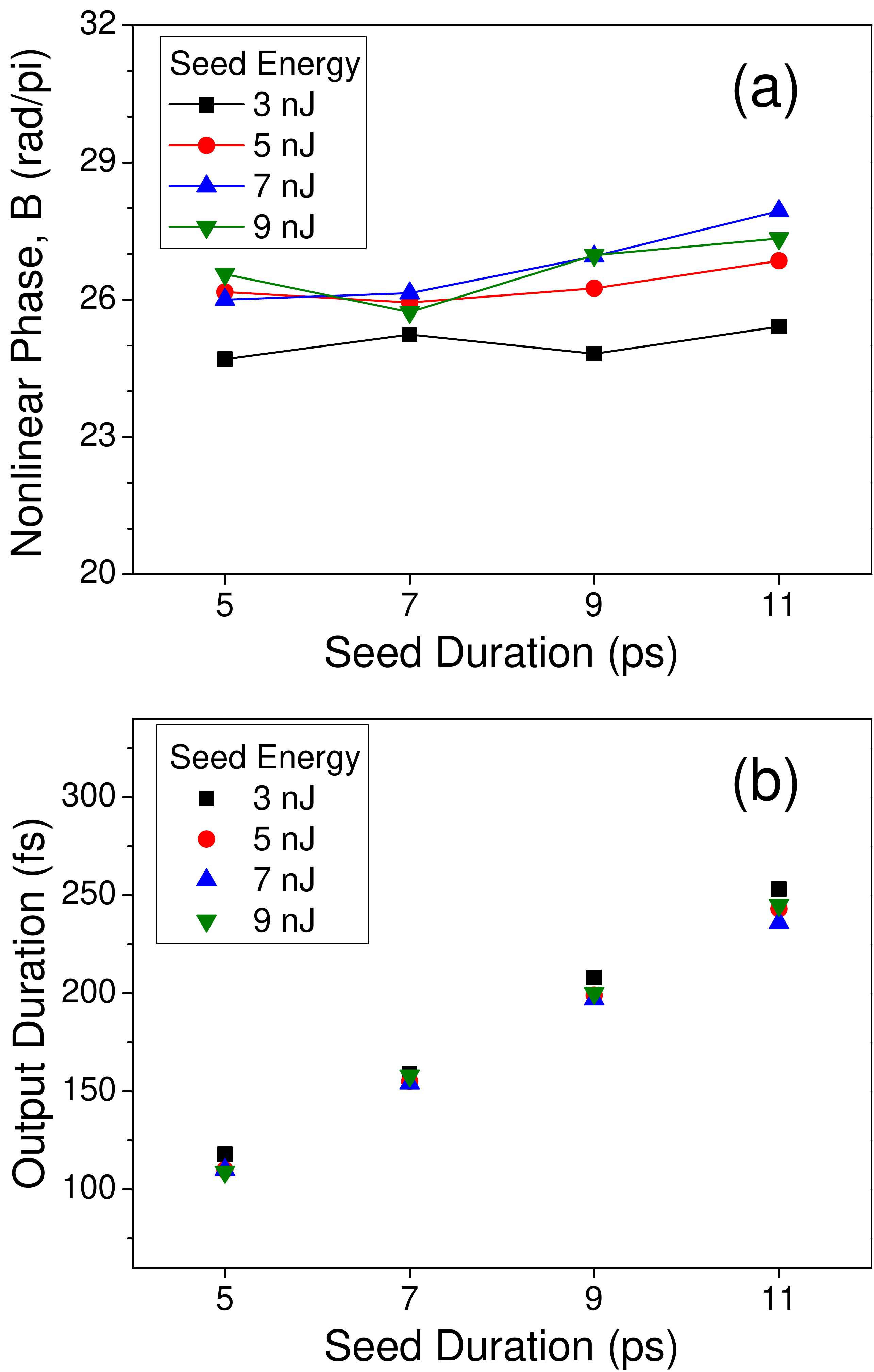}}
\caption{Numerical trends for parabolic pre-shaping, depicting (a) the total nonlinear phase accumulated and (b) output pulse duration under various seed pulse conditions.}
\label{fig:NumericalTrends}
\end{figure}

We assess the validity of our analytical model under more realistic conditions using numerical simulations.  A simple system is simulated: transform-limited, Gaussian seeds with a given full-width-half-max duration $T_\mathrm{seed}$ and energy $U_\mathrm{seed}$ are parabolically shaped in a stretch of passive fiber, the length of which is independently optimized for each seed condition.  The pulses are then amplified in a fixed-length gain fiber and compressed using a grating pair.  Nonlinear pulse propagation is modeled using the generalized nonlinear Schr\"odinger equation and the split-step method \cite{Agrawal2001}, including the effects of group-velocity dispersion, self-phase modulation, stimulated Raman scattering, and exponential gain where applicable.  We increase the gain until we reach the Raman limit, defined as when the Stokes wave contains 1\% of the total energy.  Although this threshold is arbitrary, it is approximately consistent with the limit observed experimentally (described below) and lets us compare different simulations equitably.  Repeating this process for a range of $T_\mathrm{seed}$ and $U_\mathrm{seed}$ values that is realistically realizable experimentally, we obtain the results summarized in Figure \ref{fig:NumericalTrends}, corresponding to amplified pulse energies of 2-6 $\mu$J.  The main features are immediately apparent.  Firstly, the total $B$ is relatively invariant under the simulated seed conditions, and displays no clear trends.  Secondly, the output pulse duration scales linearly with the seed duration, while its dependence on the seed energy is insignificant.  These trends are consistent with those predicted by the analytical model, suggesting that the highly simplified model we propose nevertheless captures the system's essential physical characteristics.  Indeed, the simulated transform-limited durations are consistently within 10\% of the analytically-predicted results with the same $B$, lending a level of quantitative validity to our model.  As a final note, we remark that the peak power varies only weakly with the output duration, indicating that shorter pulses can be generated without significant loss of peak power.

\section{Femtosecond Amplification System}

\subsection{Numerical Simulations}

\begin{figure}[htb]
\centering
{\includegraphics[width=0.7\linewidth]{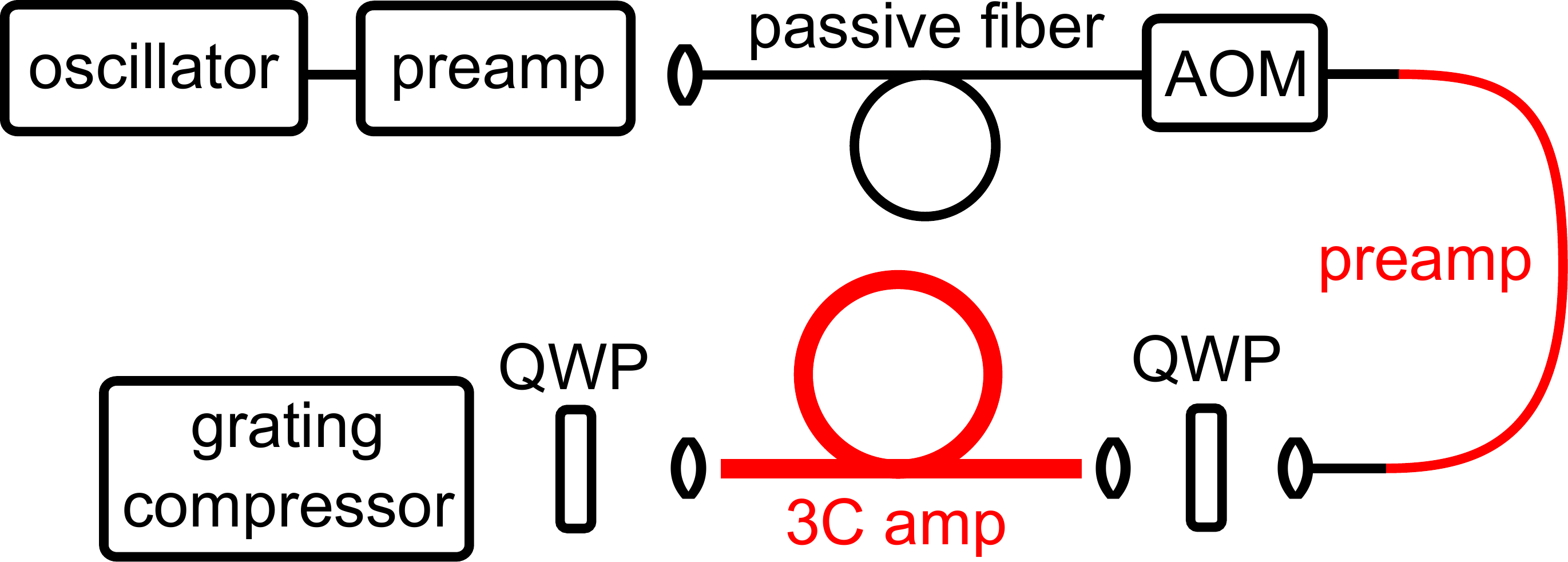}}
\caption{Schematic of experimental system.  AOM: acousto-optic modulator.  QWP: quarter-wave plate.}
\label{fig:Setup}
\end{figure}

Guided by our theoretical framework, we design a realistic femtosecond fiber amplification system based on parabolic pre-shaping as shown in Figure~\ref{fig:Setup}.  The system simulated is based closely on that experimentally realized, including realistic component pigtails and losses where applicable.  Due to availability, we take as our starting point 9 ps solitons from a commercial, Yb-doped fiber oscillator featuring robust, turnkey operation (Toptica PicoFYb).  We linearly preamplify these seed pulses prior to shaping them in order to obtain better experimental control over the shaping process; however, this would be unnecessary is a less exploratory system, since, as discussed, the final pulse duration is independent of the seed energy.  Following this step, a length of polarization-maintaining (PM), single-mode fiber (Nufern PM980-XP) nonlinearly shapes the pulses into parabolas, and they are subsequently pulse-picked by an acousto-optic modulator (Gauss Lasers AOM-M-200) and preamplified in a short, highly-doped, single-mode fiber (Nufern PM-YSF-HI).  Up to this point, the system is entirely constructed out of PM fiber, rendering it highly resistant to environmental perturbations.  The final amplification stage is a chirally coupled core (3C\textsuperscript{\textregistered}) fiber with a 34 $\mu$m core \cite{McComb2014}.  By selectively stripping away higher-order modes, this type of fiber improves single-mode behavior and reliability at high powers, which is critical for many applications.  Because 3C fiber is not PM, we use a quarter-wave plate (QWP) to circularly polarize the beam before it enters the amplifier to suppress nonlinear polarization evolution (NPE) and polarization modulation instability (PMI).  This has the additional effect of reducing the effective nonlinearity and increasing the attainable pulse energy \cite{Schimpf2009}.  Another QWP restores the linear polarization at the amplifier output.  Although our system contains free-space sections, the design can be fiber integrated for additional stability and robustness.

\begin{figure}[!t]
\centering
{\includegraphics[width=0.5\linewidth]{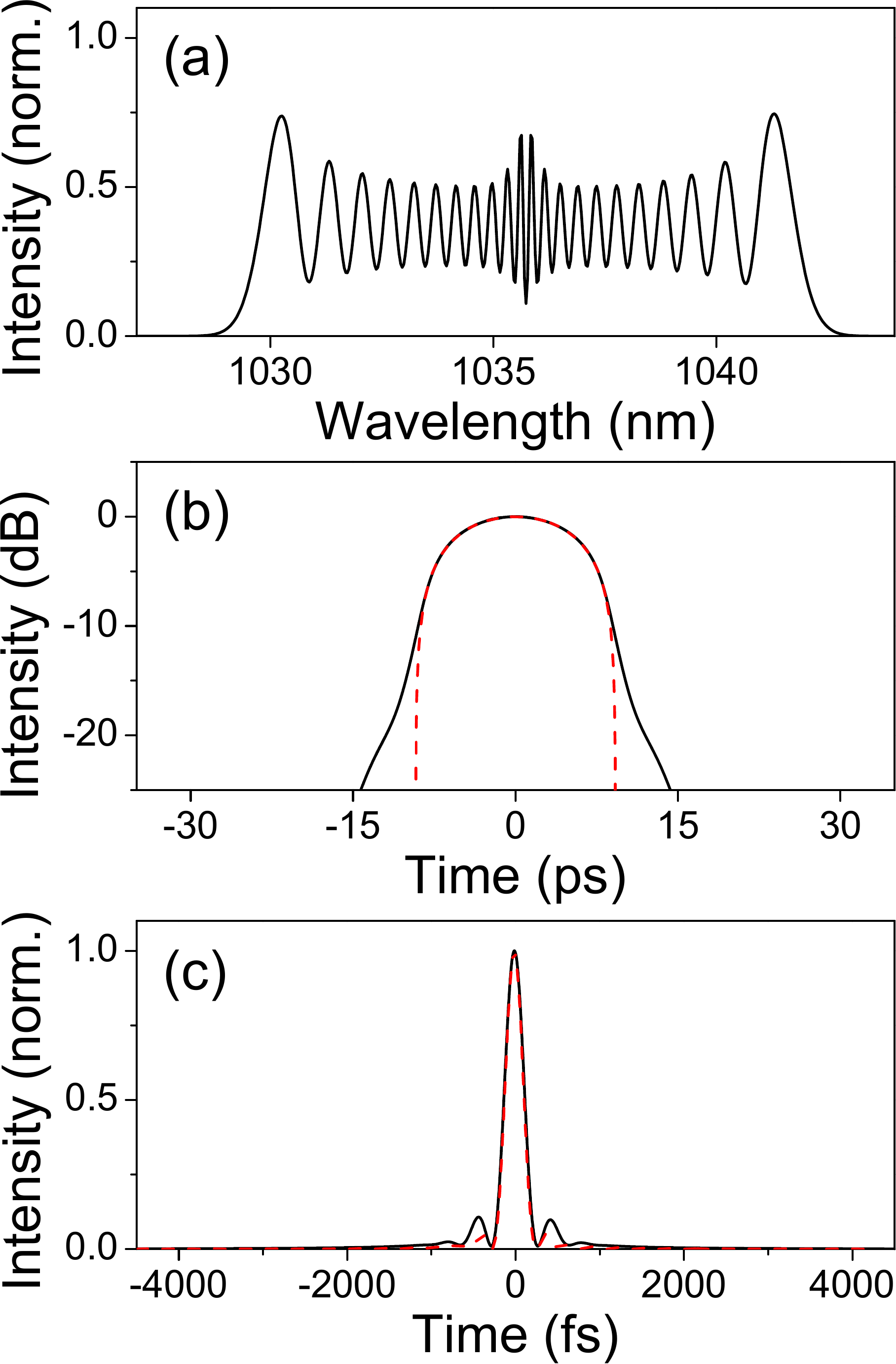}}
\caption{Simulated (a) spectral intensity; (b) chirped output (solid black) and parabolic fit (dashed red); and (c) dechirped output (solid black) and transform limit (dashed red).}
\label{fig:Sim_results}
\end{figure}

In practice, using somewhat larger seed energies reduces the burden on the final amplification stages, making it easier to access the microjoule regime.  However, excessive amplification before the shaper not only presents technical difficulties, but also increases Raman scattering in the shaper fiber, accelerating the onset of the Raman limit.  The use of numerical modeling allows us to easily optimize this balance.  Accounting for these factors, we find that using 8.2 nJ seeds and an 8.6 m passive shaping fiber strikes a favorable balance, enabling us to simulate 6 $\mu$J pulses at the amplifier output that dechirp near the 230-fs transform limit before Raman becomes noticeable (Fig.~\ref{fig:Sim_results}).

\subsection{Experimental Results}

\begin{figure}[!t]
\centering
{\includegraphics[width=0.5\linewidth]{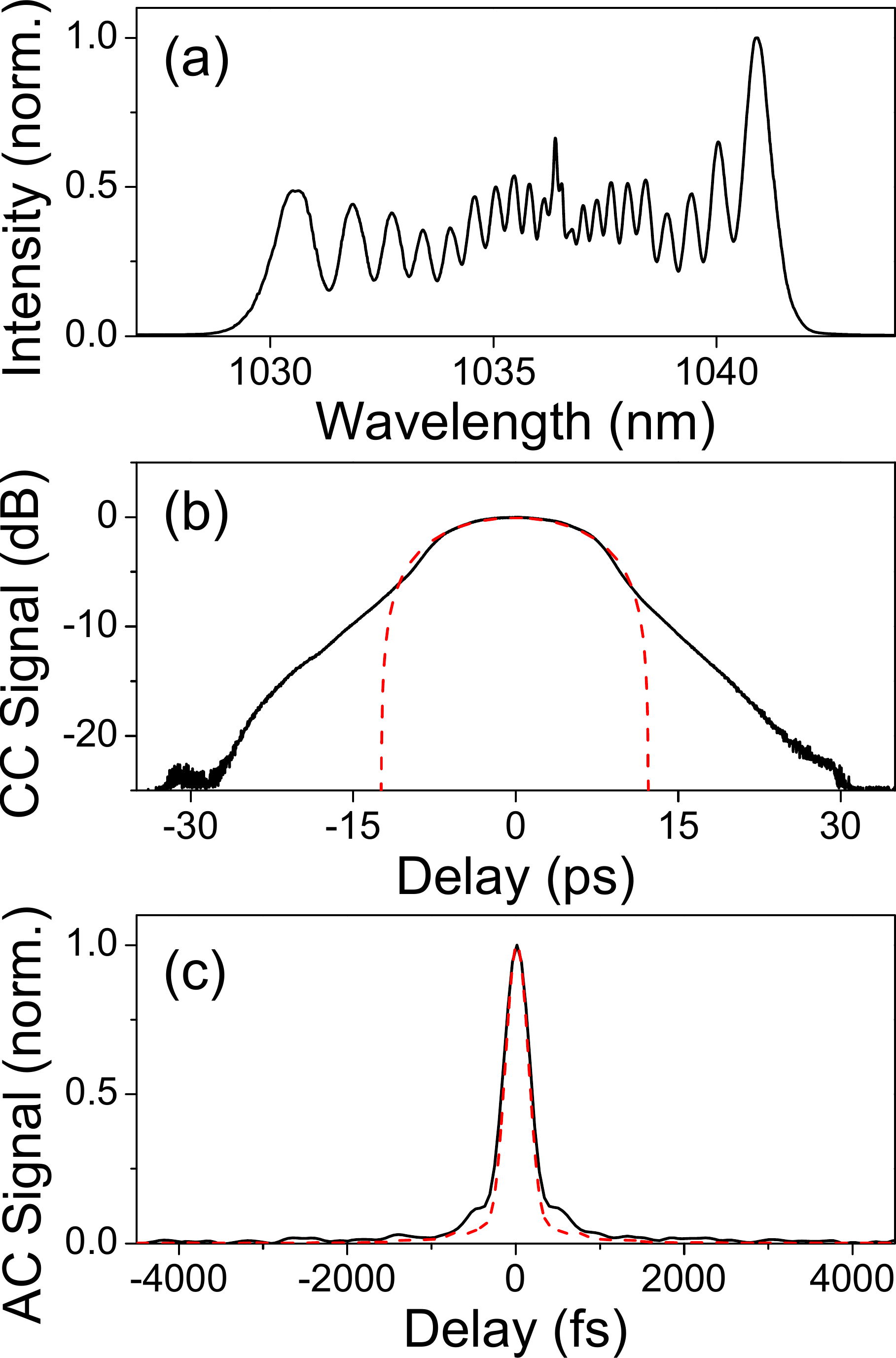}}
\caption{Experimental (a) spectral intensity; (b) cross-correlation of chirped output (solid black) and parabolic fit (dashed red); and (c) dechirped autocorrelation (solid black) and transform limited autocorrelation (dashed red).}
\label{fig:Exp_results}
\end{figure}

We assess these results experimentally by constructing the system as described above, with some additional notes given here.  The sub-nanojoule solitons from the seed oscillator are linearly preamplified to 12 nJ, of which 8.2 nJ are launched into the passive shaping fiber.  An AOM reduces the repetition rate to 576 kHz before the pulses are preamplified to 25 nJ.  Finally, the pulses are amplified in a 3C fiber counter-pumped with up to 23 W from a 976 nm, multimode diode, and are compressed using a pair of 1000 line/mm transmission gratings.  Figure~\ref{fig:Exp_results} depicts the autocorrelation and the spectrum for the best result measured, as well as a typical cross-correlation of the chirped output.  The amplified pulses are 6.1 $\mu$J, of which 4.3 $\mu$J remains after compression to 275 fs (within 13\% of the transform limit) for a compression ratio of 33.  Although a low-intensity pedestal is visible, over 85\% of the energy lies within the main peak, and launching the pulse into fiber produces the expected spectral broadening for a pulse with the measured characteristics.

We estimate from the observed spectral fringes that, consistent with numerical results, the pulse accumulates $B = 22\pi$, 12$\pi$ of which occurs in the final amplifier.  Despite these strong nonlinear effects, the pulse quality is undiminished.  Although the amplifier is non-PM, we do not observe significant NPE or PMI, and the polarization extinction ratio at the system output is typically around 13 dB.  As expected, the use of 3C fiber preserves the spatial qualities of the beam, leading to a measured beam quality factor of $\mathrm{M}^2<1.1$ along each axis.  Monitoring the cross-correlation as the pump power is varied reveals negligible change in the pulse shape and duration, affirming that the system is operating in the parabolic pre-shaping regime.  It is worth remarking that, despite the crudeness of our analytical model, it agrees well with these results, predicting an output duration of $\approx$250 fs from the measured shaped pulse duration ($\approx$18 ps) and the inferred $B$ and $B_0$.  We further note that although we lack Pierrot and Salin's full experimental details, a rough estimate using our model predicts 940 fs pulses from their system, in reasonable agreement with their observed 780 fs.

\section{Discussion}

Our experimental results match the numerical simulations closely overall, as can be seen from comparing Figures \ref{fig:Sim_results} and \ref{fig:Exp_results}.  One exception is the chirped pulse shape: the experimental cross-correlation (Fig.~\ref{fig:Exp_results}b) is less parabolic than its simulated counterpart (Fig.~\ref{fig:Sim_results}b), retaining its exponentially-decaying wings to a noticeably greater extent.  We attribute this to the shaper input deviating from the transform limit with some phase that is not well-modeled.  The discrepancy is not limiting: the clean pulse compression we observe attests to the linearization of the large nonlinear phase, and evinces a reasonably parabolic pulse shape.

\begin{figure}[htb]
\centering
{\includegraphics[width=0.5\linewidth]{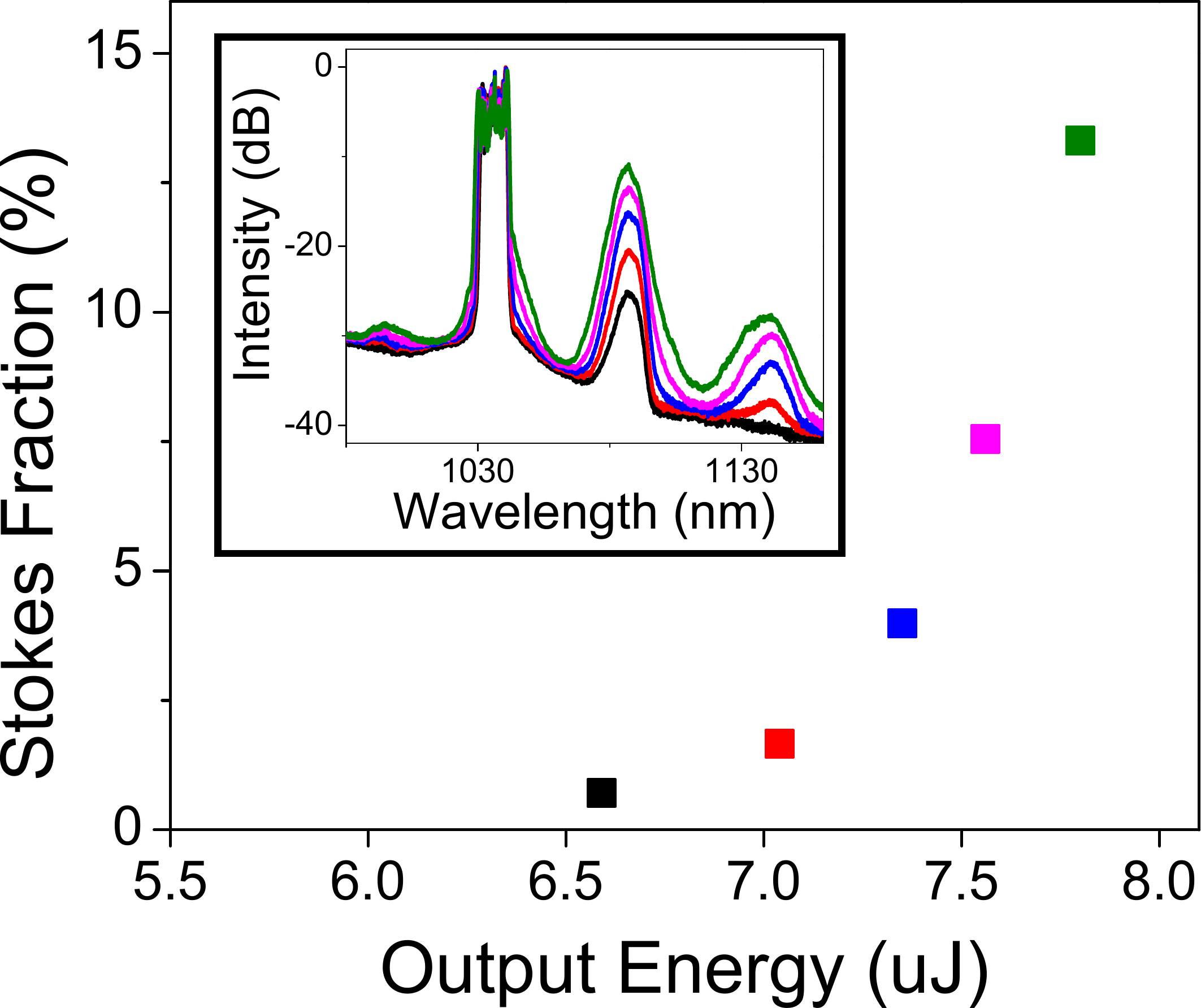}}
\caption{Stokes wave as a fraction of the total energy at different amplification levels.  Inset: corresponding spectra.}
\label{fig:Raman}
\end{figure}

As anticipated, further improvements are thwarted by Raman scattering and the growth of an incoherent Stokes wave.  Figure~\ref{fig:Raman} illustrates the measured growth of the first-order Stokes wave with rising output energy.  Experimentally, this phenomenon is accompanied by phase distortions and a loss of pulse quality when the fractional Stokes energy reaches on the order of one per cent of the primary pulse.  We speculate that this sensitivity to Raman-induced distortions is a consequence of the pulse evolution.  Raman scattering preferentially occurs near the peak of the signal pulse; thus, a Stokes wave containing a small fraction of the total energy might represent a significant local loss of intensity near the pulse center, where the parabolic profile is most critical.  The pulse shape is altered, the nonlinear phase stops being linearized, and the pulse can no longer be compressed to the transform limit.  These observations are in agreement with analytical and numerical predictions, and confirm that Raman scattering indeed fundamentally limits parabolic pre-shaping.

We end by describing how parabolic pre-shaping might be extended even further.  Our system operates at watt-level average power, as is desirable for many imaging applications; however, the average power can of course be scaled up via the repetition rate, as is a typical advantage of fiber systems.  Higher energies can be achieved simply by scaling up the amplifier's core size and reducing its length at constant $B$.  For instance, replacing the final gain stage with a short, rod-type amplifier could increase the pulse energy past $20~\mu\mathrm{J}$ at the cost of having a less practical system, as evidenced by Pierrot and Salin's original system.  In the temporal domain, additional gains could be made by further scaling equation \ref{eq:TL}.  We have shown that the Raman limit for $B$ is readily attainable in experiments; however, certain 3C fiber designs have demonstrated the ability to suppress Raman scattering using spectrally-dependent loss, and advances in this area could present a new means of improving parabolic pre-shaping systems \cite{Ma2011,Hu2012}.  Restricting ourselves to more mature technology, still shorter pulses can be obtained through additional decreases in the seed duration.  Simulations involving 3 ps seeds predict $\approx$100 fs outputs, although the maximally-attainable energy is in the neighborhood of 1-2 $\mu \mathrm{J}$.  We attribute this drop in pulse energy to the $B$ limit: since $B$ is a function of peak power, decreasing the chirped-pulse duration at constant $B$ requires a corresponding decrease in pulse energy.  The trade-offs between duration, energy, and practicality must be weighed when designing a parabolic pre-shaping system for a given application.

\section{Conclusions}

In summary, we have explored the limits of parabolic-preshaping for femtosecond pulse amplification.  We present a simplified analytical framework for describing parabolic pre-shaping, and validate it both numerically and experimentally.  Based on these results, we demonstrate an amplifier that uses parabolic pre-shaping to reach the sub-300-fs regime.  Our system compresses narrowband seed pulses by a factor of 33 and achieves nearly transform-limited, 275 fs, 4.3 $\mu$J pulses.  This combination of parameters is not only attractive for many applications, but also difficult to reach using other stretcher-free amplification techniques.  Finally, we show that Raman scattering fundamentally limits systems based on parabolic pre-shaping, and suggest routes by which such systems can be improved or scaled for various applications.

\section{Acknowledgements}

Portions of this work were supported by the National Science Foundation (ECCS-1306035) and National Institutes of Health (EB002019).  W. F. acknowledges that this material is based upon work supported by the National Science Foundation Graduate Research Fellowship Program under Grant No. DGE-1650441.

% Bibliography

\end{document}